# Academic Vibe Coding: Opportunities for Accelerating Research in an Era of Resource Constraint

**Authors:**

Matthew G. Crowson MD, MPA, MASc MBI [1,2]
Leo Celi A. Celi MD, MPH MSc [3-5]

1 Department of Otolaryngology-Head & Neck Surgery, Massachusetts Eye & Ear, Boston, Massachusetts, USA
2 Department of Otolaryngology-Head & Neck Surgery, Harvard Medical School, Massachusetts, USA
*3* Beth Israel Deaconess Medical Center, Division of Pulmonary, Critical Care and Sleep Medicine, Boston, MA
*4* Massachusetts Institute of Technology, Laboratory for Computational Physiology, Boston, MA
5 Harvard T.H. Chan School of Public Health, Department of Biostatistics, Boston, MA

**Word Count**:  3,022.

**Manuscript Submission**:

(1) Each of the authors indicated above have contributed to, read and approved this manuscript.
(2) **FINANCIAL DISCLOSURE**: no authors have disclosures related to this manuscript.
(3) **CONFLICT DISCLOSURE**: no authors have conflicts related to this manuscript.
(4) In consideration of the journal reviewing and editing my submission, the authors undersigned transfers, assigns and otherwise conveys all copyright ownership in the event that such work is published.

**Corresponding Author:**
Matthew G. Crowson MD MPA MASc MBI
Massachusetts Eye & Ear
Department of Otolaryngology-Head & Neck Surgery
243 Charles Street
Boston, Massachusetts
02114 USA
matthew_crowson@meei.harvard.edu
(p) 617-573-6559
(f) 617-573-3914

**ABSTRACT**

Academic laboratories face mounting resource constraints: budgets are tightening, grant overheads are potentially being capped, and the market rate for data-science talent significantly outstrips university compensation. Vibe coding, which is structured, prompt-driven code generation with large language models (LLMs) embedded in reproducible workflows, offers one pragmatic response. It aims to compress the idea-to-analysis timeline, reduce staffing pressure on specialized data roles, and maintain rigorous, version-controlled outputs. This article defines the vibe coding concept, situates it against the current academic resourcing crisis, details a beginner-friendly toolchain for its implementation, and analyzes inherent limitations that necessitate governance and mindful application.

**INTRODUCTION**

The modern academic research environment is facing increasing financial pressures that demand innovative approaches to maintaining research productivity. The lived experience of researchers often involves confronting rising costs for specialized inputs—personnel, sophisticated equipment, and consumables that may outpace general inflation metrics or the growth rate of their institutional or departmental budgets. This discrepancy can lead to a *de facto* decline in what research funds can procure. Indeed, for over a decade, university research budgets have experienced a decline in real purchasing power, placing strain on available resources for essential personnel and infrastructure. The United States' National Institutes of Health (NIH) budget, when adjusted for biomedical research inflation, was about 1.9% lower in real terms in 2023 than in 2003.[1] Similarly, according to the National Science Foundation, higher education research & development spending saw a seven percent increase in current dollars from 2021, but this performance in constant (inflation-adjusted) dollars is -0.4% change in 2022.[2] Compounding this, proposed fiscal policies, such as potential caps on indirect-cost recovery for grants by major funding bodies like the U.S. NIH, are projected to significantly reduce institutional operational funds. For instance, an NIH cap at 15% on indirect costs could remove an estimated US $4 billion to $9 billion annually from U.S. biomedical research centers, impacting facility maintenance, administrative support, and shared equipment.[3]

Funding acquisition and continuity also present challenges. Delays in grant renewals or new funding awards can lead to substantial cuts in laboratory spending on personnel, with reductions reported up to 50% over 16 months when anticipated funding stalls.[4] This fiscal uncertainty directly affects staffing. Institutions increasingly report that hiring freezes and the inability to offer market-competitive salaries are primary barriers to recruiting and retaining crucial data engineering (DE) and data science (DS) talent. Peer-reviewed surveys of faculty corroborate this, frequently citing "lower salary offerings" as the principal obstacle in attracting qualified data scientists, whose skills are in high demand across more lucrative private sectors; for example, a 2021 survey highlighted that 17% of industry researchers reported earning over $150,000 annually, whereas only 5% of academic researchers earned above that threshold.[5] [5] Furthermore, discretionary funds in humanities and social sciences, which often support pilot studies and methodological innovation, have seen federal support for academic humanities R&D contract significantly, more than halving from 25% of expenditures in 2007 to 9% in 2019.[6] [6] Collectively, these financial and human capital pressures create a compelling motivation for workflows that can effectively "do more" with fewer Full-Time Equivalents (FTEs).

Large Language Model (LLM)-assisted programming has demonstrated considerable potential to increase throughput in software development and data analysis tasks. Controlled studies involving tools such as GitHub Copilot have reported task completion to be approximately 55.8% faster compared to unassisted methods.[7] Similarly, research indicates substantial efficiency gains when employing GPT-4-powered coding assistants for various knowledge work and Python tasks, with some studies showing users finishing tasks 25% faster and producing 40% higher-quality results.[8]

Despite mounting evidence of LLM productivity gains, rationale and guidance on turning ad-hoc "ask-ChatGPT-for-code" habits *into a* reproducible, auditable laboratory workflow are sparse. Existing work focuses on speed metrics but seldom explains workflow integration, technology dependencies or risk mitigation. This paper introduces "Vibe coding," a concept defined as structured, prompt-driven code generation with LLMs embedded in reproducible research workflows. Given the mounting resource constraints and the demonstrated productivity enhancements of LLM-assisted programming, this paper aims to address the gap between the potential of these tools and their systematic integration into academic research. We will explore how Vibe coding can offer a practical response to the current academic resourcing crisis by compressing the idea-to-analysis timeline, reducing staffing pressure on specialized data roles, and maintaining rigorous, version-controlled outputs, while also analyzing its inherent limitations that necessitate careful governance.

**WHAT IS VIBE CODING?**

Vibe coding is a methodology that leverages LLMs to convert high-level conceptual plain-language inputs, termed "vibes," into executable, version-controlled research artifacts. These vibes typically encompass a problem statement, the context of the data being analyzed, and proposed methodological approaches for analyzing that data. The conversion is achieved through structured prompt templates rather than ad-hoc, conversational interactions with an LLM. The outputs include scripts, data processing pipelines, analysis notebooks, and preliminary interpretative summaries.

The distinction between vibe coding and casual LLM interaction for code generation is important for its effective application in research (**Table 1**). The paradigm shift inherent in vibe coding is from generating small, illustrative code examples to establishing repeatable, auditable analytical pipelines that might better withstand the scrutiny of peer review and meet the requirements for replication studies. It emphasizes systematic LLM application within a controlled research environment. Vibe coding also democratizes access to coding languages and outputs as one does not necessarily need to be proficient in coding to leverage vibe coding.

**Table 1**. Differences between ad-hoc code chats and vibe coding workflows.

| Attribute | Ad-hoc "chat coding" | Vibe coding workflow |
|---|---|---|
| Interaction style | Episodic, conversational Q&A | Template-based prompts, version-controlled (e.g., git) |
| Output scope | Isolated snippets, often context-poor | Complete, runnable modules, documentation |
| Quality control | Primarily manual inspection by the user | Automated unit tests, continuous integration (CI) |
| Re-use | Low; context-dependent and perishable | Prompts and outputs stored, templated, and shareable |
| Reproducibility | Difficult; relies on chat history recall | High; defined by versioned prompts and code |

**EMBEDDING VIBE CODING IN RESEARCH WORKFLOWS**

Vibe coding can be integrated at multiple stages of the typical research lifecycle, transforming traditional hand-offs between specialized roles into structured, LLM-assisted tasks. This might alleviate bottlenecks, particularly those related to data science and engineering expertise. Discrete research tasks, which often require specialized expertise and, ergo, represent potential delays, can be reframed as inputs for structured prompt templates (**Table 2**). This approach does not eliminate the need for the expert but can make the initial drafting and iteration processes more efficient, reducing the critical path dependency on scarce DE/DS resources.

**Table 2**. Vibe coding can assist along the entire computational research workflow.

| Canonical step | Human Research Role(s) | Example vibe coding prompt template (conceptual) | LLM output (illustrative) |
|---|---|---|---|
| **Data ingestion & cleaning** | Data engineer, Research assistant | "Using Python pandas, clean the attached CSV file ([dataset_name].csv) representing hospital patient stays: 1. Drop rows where 'LengthOfStay' is negative or missing. 2. For remaining missing 'LengthOfStay' values, impute using the median 'LengthOfStay' for that patient's Diagnosis Related Group ('DRG'). 3. Ensure columns X, Y, Z are numeric. 4. Return a tidy pandas DataFrame and a summary of changes." | A Python script implementing the cleaning logic with inline comments and a brief report on imputation and data removal. |
| **Exploratory analysis** | Data scientist, Researcher | "Generate a Jupyter notebook for exploratory data analysis on the cleaned patient stay data: 1. Provide descriptive statistics for key variables (Age, LengthOfStay, NumberOfProcedures). 2. Create box-plots for 'LengthOfStay' grouped by 'DRG'. 3. Identify and list the top 1% of 'LengthOfStay' outliers. 4. Generate pairwise correlation plots for numeric variables." | An executable Jupyter notebook with code cells for each requested analysis, including visualizations and brief annotations. |
| **Statistical modelling** | Biostatistician, Researcher | "Draft an R script to fit a mixed-effects Poisson regression model. The outcome variable is 'ReadmissionCount', with a random intercept for 'HospitalID'. Fixed effects include 'Age', 'DRG_Category', and 'LengthOfStay'. Calculate robust standard errors. Provide a summary of model coefficients and a plain-language interpretation of | An R script containing model fitting code (e.g., using lme4 or glmmTMB) and a markdown section for results interpretation. |

| | | |
|---|---|---|
| | the 'LengthOfStay' effect." | |
| **Project tracking** | Project manager, Lab head | "Produce a Markdown Gantt chart for a research project starting June 1, [Year]. Phases: Data Collection (6 weeks), Data Analysis (4 weeks), Manuscript Preparation (5 weeks). Include weekly milestones for each phase." | A Markdown-formatted Gantt chart suitable for inclusion in project management tools or documentation (e.g., GitHub Projects). |
| **Reproducibility bundle** | Research assistant, Librarian | "Prepare a reproducibility bundle for archiving on the Open Science Framework (OSF). The bundle should include: all Python and R scripts in a /code directory, a requirements.txt file for Python dependencies, an R environment manifest, a Makefile to run the analysis pipeline, and a brief README.md describing the project and data (use [metadata_file].json for details)." | A structured directory with the specified files and a template README.md. |

## TOOL CHAIN: A HAND-HELD GUIDE FOR NON-TECHNOLOGISTS

Adopting vibe coding in an academic lab doesn't require extensive technical expertise. A selection of human-friendly tools can help researchers leverage vibe coding. These tools assist with guiding the models, accessing its capabilities, ensuring reliable execution of generated programs, and keeping track of all research components.

***Guiding the Models: Prompting Tools and Coding Assistants.*** Researchers typically start interacting with the model (Large Language Model (LLM)) through familiar software environments, often enhanced with AI assistants. For example, common coding software like Visual Studio Code (Microsoft, Redmon, WA, USA) can be equipped with tools like GitHub Copilot (Microsoft, Redmon, WA, USA). These assistants can offer suggestions and generate code directly as the researcher types or through a chat-style conversation (e.g. "("Load my CSV and plot mean blood pressure by age bracket"), allowing for a back-and-forth process to refine instructions given to the LLM. Some newer code editors, such as Cursor (Cursor, San Francisco, CA, USA), are specifically built around this idea of working closely with an AI partner. For researchers who analyze data using tools like JupyterLab (Linux Foundation Charities, San Francisco, CA, USA) specialized extensions can simplify how they create and manage their instructions for the LLM. These add-ons handle many of the technical details behind the scenes, so researchers can focus on clearly describing their research needs and data to the AI. The common point is ease: researchers stay inside a single window, refining questions and seeing immediate, runnable answers.

***Accessing LLM Models.*** There are different ways to access the LLM models that power vibe coding, depending on the lab's needs, emphasizing data privacy and control. Companies like OpenAI (OpenAI Inc., San Francisco, CA, USA; known for Generative Pre-trained Transformer "GPT" models), Anthropic (Anthropic AI, San Francisco, CA, USA; the 'Claude' models), and Google (Alphabet Inc., Mountain View, California, CA; 'Gemini' models) offer powerful LLMs that can be accessed via hosted public portals and application programming interface (APIs). Labs usually pay based on how much they use the service, and they can often start with free access or small budgets to see how useful and costly these models are for their specific research.

For research involving sensitive information such as personally identifiable information (PII) and/or protected health information (PHI) or when a lab needs full control over how the LLM is run, it's possible to use "open weight" models that can run on institution hardware. These are LLMs, such as Llama-3 (Meta Inc., Menlo Park, CA) or Mixtral (Mistral AI, Paris, France), that can be installed and run on a lab's on-premises computers, typically requiring graphics processing units (GPUs). This ensures sensitive data stays within the institutions' secure environment. Some institutions with more AI maturity, however, have private access to the proprietary closed models (e.g., private instance of OpenAI models).

***Ensuring Reliable and Repeatable Analysis.*** A central goal in research is ensuring that analyses are reliable and can be repeated by others. Containerization" tools, like Docker or Podman, help achieve this by packaging all the necessary software for an analysis into a self-contained "bundle." This means that any computer code generated by the LLM will run the same way every time, on any computer, because it explicitly defines the software environment dependencies to run that code. Even researchers who aren't comfortable with complex computer commands can usually start these bundles easily.

To further ensure the quality of the AI-generated programs, automated checking systems can be used. Tools often associated with "Continuous Integration" (CI; like GitHub Actions or GitLab CI/CD) can automatically run tests on any new or changed program. These systems can check if the program works correctly, follows good coding practices, and even help prepare any accompanying documentation. This helps make sure that the results of vibe coding are dependable.

***Keeping Track of Research Components and Pipelines.*** Careful record-keeping is vital in research. "Version control" systems, with Git being the most common, are essential for tracking changes to all parts of a project, not just the computer code, but also the specific instructions (prompts) given to the LLM. These prompt files should be saved alongside the code they helped create. This allows anyone reviewing the research to understand exactly how each piece of analysis was generated.

When research involves very large datasets or complex analytical steps that are too big for standard version control, tools like Data Version Control (DVC) can help. DVC works with Git to keep track of these large elements without cluttering the main project record.  Finally, to make the vibe coding workflow easier to use, simple custom commands can be created. These commands can bundle a series of steps into one straightforward instruction (AKA 'pipelines'), hiding the complexity and making it easier for everyone in the lab to use the workflow consistently.

A lab can typically start trying out this set of tools on a reasonably modern computer with about 32 GB of memory (RAM). Powerful graphics cards (GPUs) are not needed when calling LLMs over the internet/APIs, but they become important if the lab decides to run large open-weight models on its computers or wants to train or fine-tune AI models.

**LIMITATIONS AND RISKS**

While vibe coding offers significant potential for accelerating research, it introduces non-trivial risks that require careful management and governance (**Table 3**). Ignoring these risks can undermine the validity and integrity of the research.

*Data Privacy and Security*. Transmitting sensitive or protected data (e.g., Protected Health Information (PHI), personally identifiable information) to third-party proprietary LLM APIs introduces risks of data breaches or residual memory persistence at the provider. Researchers can mitigate this risk through employing retrieval-augmented generation (RAG) with locally cached and redacted data, utilize services with strict data privacy agreements (e.g., HIPAA-compliant offerings if available); or, for maximum security, run open-weight models entirely on-premises behind institutional firewalls.

*Code Correctness and Reliability*. LLMs can "hallucinate" non-existent variables or functions, apply statistically inappropriate tests for the given data type or research question, or silently omit crucial edge-case handling in generated code. Operators can mandate the generation of unit tests alongside functional code. It is also a best practice to have human domain experts (e.g., statisticians, senior researchers) rigorously review all generated code and its outputs, particularly for methodologies and interpretations, before any results are considered for publication or decision-making.

*Licensing and Intellectual Ownership*. Code generated by LLMs may inadvertently embed or be derived from snippets of code from the model's vast training corpus, which can include various software licenses (e.g., GPL, MIT). As a result. the intellectual property status of LLM-generated code can be ambiguous. Researchers should incorporate Software Package Data Exchange (SPDX) headers or similar licensing declarations in all generated script files. Standard research bookkeeping tasks like recording the prompts used, the LLM version, and cryptographic hashes of the generated outputs should be completed for auditing purposes. Researchers should also consult institutional legal counsel regarding IP policies for AI-generated content.

***Cost Creep and Resource Management.*** While individual API calls for token generation may seem trivial in cost (i.e., cents per token), frequent and iterative prompting, especially with large contexts or complex requests using the most powerful models, can lead to escalating operational expenses. Researchers should consider implement caching mechanisms for LLM outputs to avoid re-generating identical results. Optimizing prompt structures for conciseness and clarity can also decrease token volumes. If appropriate for the task, defaulting to using smaller, less expensive models for routine or less complex tasks, reserving larger models for tasks demanding higher reasoning capabilities. Set and monitor API usage budgets.

***Skill Dilution and Over-Reliance.*** Continuous reliance on LLM-generated code, particularly among trainees or researchers less experienced in programming, could potentially erode foundational coding proficiency and critical analytical thinking skills. Research teams should counterbalance LLM use with structured code review sessions where trainees explain the generated code. Educators can incorporate "manual-only" coding exercises or project components into training programs. Above all else, research leadership should emphasize that LLMs are tools for assistance, not replacements for understanding fundamental principles.

**Table 3**. Common vibe coding risks and their mitigation strategies.

| Category | Risk | Primary Mitigation Strategies |
|---|---|---|
| **Privacy & Security** | PHI/PII leakage via external API calls | On-premises open-weight models; RAG with redacted data; strict data minimization; data use agreements with API providers. |
| **Correctness** | Statistical errors, logical flaws, bugs | Mandatory unit tests executed via CI; rigorous senior researcher and domain expert review; code walkthroughs. |

| Cost | Escalating API subscription/token spend | Output caching; prompt compression/optimization; tiered model usage (smaller models for simpler tasks); budget monitoring. |
| Licensing & IP | Unclear provenance, license non-compliance | SPDX headers; comprehensive logging of prompts, LLM versions, and output hashes; consultation with legal/IP offices. |
| Skills | Atrophy of coding depth, critical assessment | Scheduled manual coding sprints; peer code reviews focused on understanding; clear guidelines on LLM as an assistive tool. |

## CONCLUSION

Vibe coding presents a structured approach to converting the recognized productivity gains of LLMs into a coherent and auditable research workflow. This methodology directly addresses two significant structural challenges prevalent in contemporary academic research: progressively constrained budgets and the pronounced scarcity of specialist data science and engineering talent. By elevating prompts to the status of first-class research objects, which are versioned, meticulously reviewed, and integral to reproducible analyses, research laboratories can potentially accelerate the pace of discovery without compromising the foundational tenets of scholarly rigor.

The paradigm of vibe coding does not advocate for the abolition of human expertise. Instead, it proposes a strategic reallocation of scarce human data science and data engineering hours away from automatable or draft-level coding tasks towards aspects of research where human judgment, critical thinking, and nuanced interpretation are irreplaceable. The limitations surrounding data privacy, code correctness, operational cost, and intellectual property demand robust governance frameworks and mindful implementation strategies. In no uncertain terms, vibe coding is no replacement for human-in-the-loop expert guidance and oversight. However, these challenges are not inherently disqualifying.

When implemented thoughtfully within a framework that includes containerization for reproducibility, continuous integration for quality assurance, and transparent management of prompts and generated artifacts, vibe coding stands as a viable, largely vendor-neutral accelerator for modern scholarship. It offers a pragmatic pathway for academic researchers, including those without extensive programming backgrounds, to leverage cutting-edge AI capabilities in their pursuit of knowledge.

**ACKNOWLEDGEMENTS**

Nil.

**TABLES**

**Table 1.**   Differences between ad-hoc code chats and vibe coding workflows.

**Table 2.**   Vibe coding can assist along the entire computational research workflow.

**Table 3.** Common vibe coding risks and their mitigation strategies.